\documentclass[aps,twocolumn,superscriptaddress,footinbib]{revtex4-1}

\usepackage{amsmath}    
\usepackage{color}
\usepackage{graphicx}   
\usepackage{verbatim}   
\usepackage{color}      
\usepackage{subfigure}  
\usepackage{hyperref}
\usepackage{units} 
\usepackage{blindtext}

\begin{document}

\title{Effect of lattice excitations on transient near edge X-ray absorption spectroscopy}

\author{N. Rothenbach}\email[] {nico.rothenbach@uni-due.de}
\affiliation{Faculty of Physics and Center for Nanointegration Duisburg-Essen (CENIDE), University of Duisburg-Essen, Lotharstr.~1, 47057 Duisburg, Germany}

\author{M. E. Gruner}
\affiliation{Faculty of Physics and Center for Nanointegration Duisburg-Essen (CENIDE), University of Duisburg-Essen, Lotharstr.~1, 47057 Duisburg, Germany}

\author{K. Ollefs}
\affiliation{Faculty of Physics and Center for Nanointegration Duisburg-Essen (CENIDE), University of Duisburg-Essen, Lotharstr.~1, 47057 Duisburg, Germany}

\author{C. Schmitz-Antoniak}
\affiliation{Peter-Gr\"{u}nberg-Institut (PGI-6), Forschungszentrum J\"{u}lich, 52425 J\"{u}lich, Germany}

\author{S. Salamon}
\affiliation{Faculty of Physics and Center for Nanointegration Duisburg-Essen (CENIDE), University of Duisburg-Essen, Lotharstr.~1, 47057 Duisburg, Germany}

\author{P. Zhou}
\affiliation{Faculty of Physics and Center for Nanointegration Duisburg-Essen (CENIDE), University of Duisburg-Essen, Lotharstr.~1, 47057 Duisburg, Germany}

\author{R.~Li}
\altaffiliation{currently at Department of Engineering Physics, Tsinghua University, Beijing 100084, China}
\affiliation{SLAC National Accelerator Laboratory, 2575 Sand Hill Rd., Menlo Park, California 94025, USA}

\author{M. Mo}
\affiliation{SLAC National Accelerator Laboratory, 2575 Sand Hill Rd., Menlo Park, California 94025, USA}

\author{S. Park}
\altaffiliation{currently at Brookhaven National Laboratory, Center for Functional Nanomaterials, Upton, NY 11973-5000, USA}
\affiliation{SLAC National Accelerator Laboratory, 2575 Sand Hill Rd., Menlo Park, California 94025, USA}

\author{X. Shen}
\affiliation{SLAC National Accelerator Laboratory, 2575 Sand Hill Rd., Menlo Park, California 94025, USA}

\author{S. Weathersby}
\affiliation{SLAC National Accelerator Laboratory, 2575 Sand Hill Rd., Menlo Park, California 94025, USA}

\author{J. Yang}
\altaffiliation{currently at Center of Basic Molecular Science, Department of Chemistry, Tsinghua University, Beijing 100084, China}
\affiliation{SLAC National Accelerator Laboratory, 2575 Sand Hill Rd., Menlo Park, California 94025, USA}

\author{X. J. Wang}
\affiliation{SLAC National Accelerator Laboratory, 2575 Sand Hill Rd., Menlo Park, California 94025, USA}

\author{O. \v{S}ipr}
\affiliation{FZU Institute of Physics of the Czech Academy of Sciences, Cukrovarnicka 10, CZ-162 53 Prague, Czech Republic}

\author{H.~Ebert}
\affiliation{Universit\"at M\"unchen, Department Chemie, Physikalische Chemie, Butenandtstr. 5-13, 81377 M\"unchen}

\author{K. Sokolowski-Tinten}
\affiliation{Faculty of Physics and Center for Nanointegration Duisburg-Essen (CENIDE), University of Duisburg-Essen, Lotharstr.~1, 47057 Duisburg, Germany}

\author{R. Pentcheva}
\affiliation{Faculty of Physics and Center for Nanointegration Duisburg-Essen (CENIDE), University of Duisburg-Essen, Lotharstr.~1, 47057 Duisburg, Germany}

\author{U. Bovensiepen}
\affiliation{Faculty of Physics and Center for Nanointegration Duisburg-Essen (CENIDE), University of Duisburg-Essen,
Lotharstr.~1, 47057 Duisburg, Germany}

\author{A. Eschenlohr}
\affiliation{Faculty of Physics and Center for Nanointegration Duisburg-Essen (CENIDE), University of Duisburg-Essen,
Lotharstr.~1, 47057 Duisburg, Germany}

\author{H. Wende}\email[] {heiko.wende@uni-due.de}
\affiliation{Faculty of Physics and Center for Nanointegration Duisburg-Essen (CENIDE), University of Duisburg-Essen, Lotharstr.~1, 47057 Duisburg, Germany}

\date{\today}

\begin{abstract}

Time-dependent and constituent-specific spectral changes in soft near edge X-ray spectroscopy (XAS) of an [Fe/MgO]$_8$ metal/insulator heterostructure upon laser excitation are analyzed at the O K-edge with picosecond time resolution. The oxygen absorption edge of the insulator features a uniform intensity decrease of the fine structure at elevated phononic temperatures, which can be quantified by a simple simulation and fitting procedure presented here. Combining X-ray absorption spectroscopy  with ultrafast electron diffraction measurements and ab initio calculations demonstrate that the transient intensity changes in XAS can be assigned to a transient lattice temperature. Thus, the sensitivity of transient near edge XAS to phonons is demonstrated.

\end{abstract}

\maketitle

\section{Introduction}
X-ray absorption spectroscopy (XAS) is a versatile tool to analyze the local electronic structure, the magnetic properties or the chemical environment of condensed matter in an element- and orbital-sensitive way. With the continuous development of X-ray sources, especially synchrotron radiation sources, XAS became one of the most powerful techniques to investigate material properties in many different fields of science, including molecular and atomic physics, cell biology, life science, catalysis as well as nanotechnology \cite{Mobilio_Springer_15}. Individual spectroscopic signatures in static X-ray absorption spectra have been proven to contain crucial information about the material properties of the investigated systems, ranging from structural information, like the bonding character of water in the bulk liquid phase \cite{Wernet_Science_04}, to information about the magnetic properties  \cite{Gambardella_Science_03} and coupling mechanism between molecules and ferromagnetic films \cite{Wende_NatMat_07}. The importance of the analysis of X-ray absorption spectra in the soft X-ray regime at the oxygen K-edge has been recently reviewed for a broad variety of systems ranging from molecular systems to solid oxides by Frati et al. \cite{Groot_review_20}. The analysis of the oxygen K-edge provides deep insight e.g. into the bonding and the local surrounding of the oxygen atoms by studying the detailed fine structures of the XAS.\\
The availability of high-quality femtosecond (fs) X-ray sources like synchrotron femtoslicing sources \cite{Holldack_JoSR_14} and X-ray free electron lasers \cite{Ackermann_NatPhot_07,Emma_NatPhot_10,Schlotter_RevSciInstr_12} gave rise to a vastly increasing number of studies of dynamic processes such as photosynthesis \cite{Wolf_NatComm_17} or photocatalysis \cite{Wernet_Nature_15} and characterization of non-equilibrium states in condensed matter induced by (optical) laser excitation. However, the ability to truly extend the X-ray spectroscopy to the ultrafast timescale depends on a deeper understanding of the non-equilibrium situation  \cite{Chergui_review,Pertot_XAS}. For instance, while static X-ray spectroscopy is commonly used to disentangle the spin and orbital contribution to the total magnetic moment, the validity of the X-ray magnetic circular dichroism (XMCD) sum rules had first to be proven for highly non-equilibrium states \cite{Carva_EPL_09}. Time resolved studies at the oxygen K-edge have been performed e.g. on VO$_2$ to analyze the photoinduced insulator-metal phase transition in Ref. \cite{Cavalleri_PRL_05} as well as molecular systems \cite{Beye_PRL_2013,Wolf_NatComm_2017} and atomic oxygen on surfaces \cite{Beye_JPCL_2016} to analyze the bonding modifications upon optical excitation. In these studies on femtosecond timescales often very specific XAS fine structures are analyzed in narrow photon energy regimes of a few eV. In the present work we study for longer delay times of 90 picoseconds a much broader energy range of about 40 eV in the vicinity of the oxygen K-edge. The optical excitation and the attendant excitation and relaxation of electronic and lattice degrees of freedom influence the spectroscopic fine structures also in this broader energy regime in a way which is so far not well understood. Hence, it is necessary to systematically investigate this pump-induced spectral changes for different timescales and materials. On the time scale of 20-100 picoseconds the phonon sub-systems are equilibrated and the solid heterostructures studied in this work are close to their fully thermalized states. Therefore, it is interesting to connect the modifications of the various XAS fine structures in the transient XAS to the changes of fine structures of static XAS measurements as a function of temperature. Detailed static temperature dependent investigations have been shown recently for solid oxides \cite{Nemausat_PCCP_2017} also stressing the importance of quantum vibrations. In this work the near edge XAS of MgO was analyzed in experiment and theory at the Mg K-edge. Clear modifications of the fine structures in the Mg XAS versus temperature could be identified which were linked to changes of the local surrounding of the Mg atoms. In the present work we focus on the oxygen K-edge in Fe/MgO multilayers because of the importance of this edge in numerous systems as stressed in Ref. \cite{Groot_review_20}. This will allow for a more fundamental understanding of XAS fine structure modifications on wider photon energy ranges paving the way towards time resolved EXAFS measurements on ultrafast time scales in the future.

\begin{figure}
    \centering
    \includegraphics[width=\columnwidth]{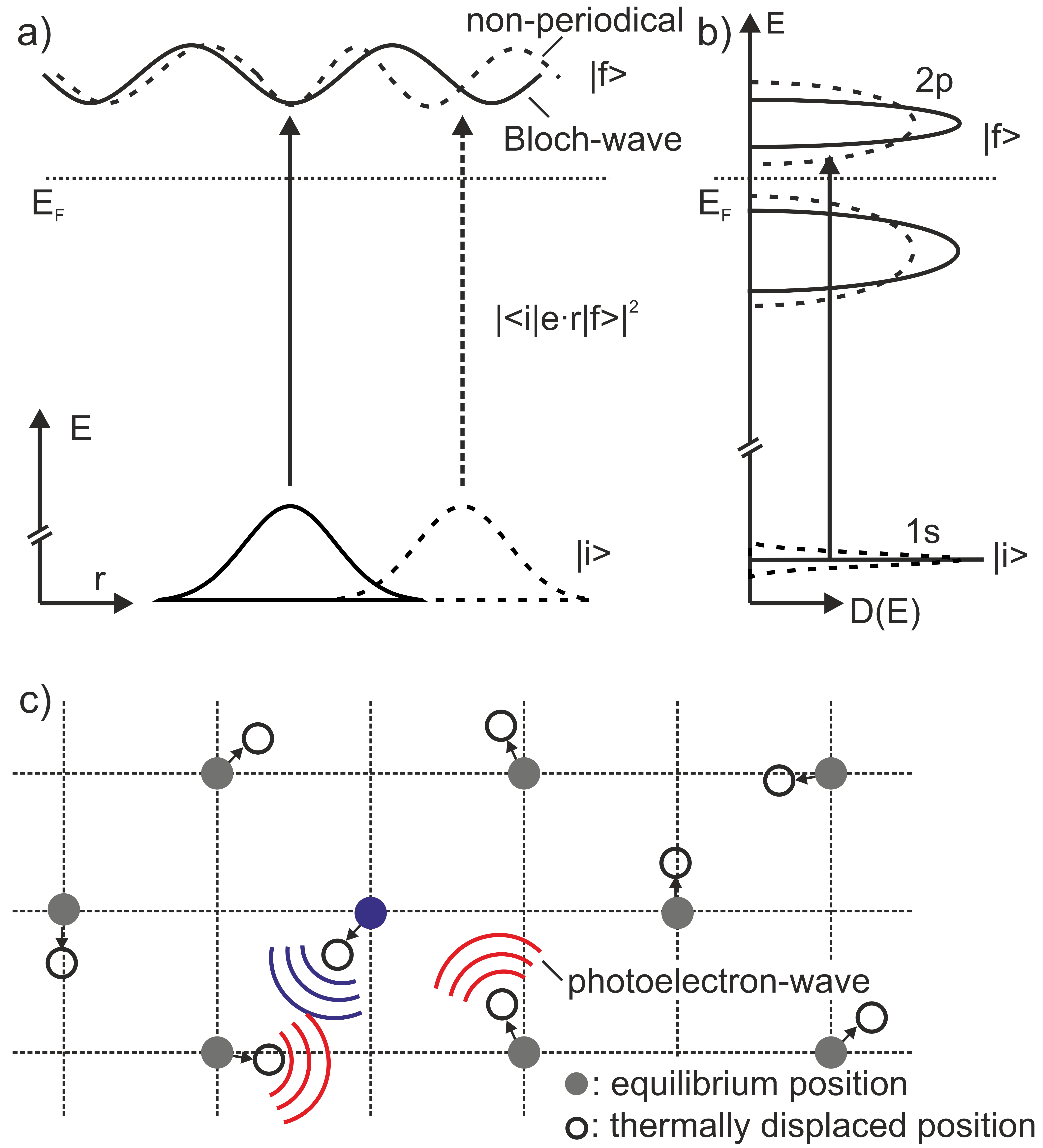}
    \caption{Schematic presentations of the X-ray absorption process including the effect of structural thermal disorder (phonons) by analyzing: a) the wave functions of the initial state $|i\rangle$ and the final state $|f\rangle$ without (solid line) and with structural thermal disorder (dashed line), b) the density of the initial and final states and c) a real-space representation of the outgoing photoelectron wave from the absorbing atom (marked blue) and the backscattered photoelectron wave from the structurally disordered surrounding atoms. The atomic equilibrium positions are depicted by full circles whereas the thermally disordered positions are shown as open circles. For illustration purpose the displacement in the initial states $|i\rangle$ is intentionally exaggerated and not to scale.}
    \label{fig:Rothenbach2_PRB}
\end{figure}

To motivate the sensitivity of X-ray absorption spectra in the near edge regime to thermal disorder (thermal atomic displacements), Fig.~\ref{fig:Rothenbach2_PRB} shows different schematic presentations of the effect of thermally excited structural disorder on the X-ray absorption process. Fig.~\ref{fig:Rothenbach2_PRB}a) describes the X-ray absorption process: we analyze the wave functions of the absorbing atom, where, for the equilibrium state (solid lines), the electron is excited from the localized initial state $|i \rangle$, presented by a simple wave function, to the periodical Bloch-wave final state $|f\rangle$. The excitation of the electron is here shown vertically (sudden approximation, Born-Oppenheimer approximation). The transition probability, given by Fermi's golden rule, is proportional in the dipole approximation to the transition matrix element  $|\langle i | e \cdot r | f \rangle |^2$ (Fig.~\ref{fig:Rothenbach2_PRB}a). While the effect of thermal disorder (dashed lines), i.e. structural distortions, leads to a dislocation of the localized initial state without influencing the shape of the wave function, the final state is strongly modified by the thermal fluctuations of the surrounding atoms, changing the Bloch-wave locally to a non-periodic function. In the density of states picture, schematically shown in Fig.~\ref{fig:Rothenbach2_PRB}b) for the case of the oxygen K absorption edge, i.e. investigating the transition from the localized $1s$ initial state to the  $2p$ final states (dipole approximation), the effect of thermal disorder (dashed lines) can be described by a broadening of both states, which is stronger for the more delocalized $2p$ final states. Lastly, as depicted in Fig.~\ref{fig:Rothenbach2_PRB}c), the X-ray absorption process and more precisely the origin of the detailed fine structures of the near edge X-ray absorption spectra can be described in a multiple-scattering approach \cite{Ankudinov_PRB_98}. The final state originates from the interference of the outgoing photoelectron wave and the multiple scattered wave at the surrounding atoms. Since the surrounding atoms are displaced from their equilibrium positions by the thermal disorder (open circles) also relative to the photo-excited atom (blue circle), the multiple scattering of the photoelectron wave, and consequently the final state $|f\rangle$ is modified by the thermal distortion. In summary, all different viewpoints schematically presented in Fig.~\ref{fig:Rothenbach2_PRB} indicate that both, the initial state by displacement of the absorbing atom and the final state by the scattering at the surrounding atoms are modified by the thermal disorder. Consequently, the X-ray absorption coefficient is influenced also in the near edge regime by thermal disorder which effects both, the initial and the final states. Here, we analyze in which way this can be detected for transient near edge X-ray absorption spectra after excitation with a UV pulse. The thermalization of the system after some ten picoseconds is revealed by complementary ultrafast electron diffraction experiments which will provide the phonon temperature.

The Fe/MgO system investigated here has already been introduced as a model system for ultrafast energy transfer studies in metal/oxide heterostructures in our previous work \cite{RothenbachFeMgO,Gruner_PRB_19}. By a $50$\,fs, $266$\,nm laser pump pulse, the electronic system of the Fe constituent is excited exclusively. The highly non-equilibrium Fe electronic subsystem thermalizes within the first hundreds of femtoseconds by electron-electron scattering, and then redistributes the initial input energy via phononic processes among the heterostructure. An ultrafast energy transfer across the Fe-MgO interface has been identified which is mediated by high-frequency, interface vibrational modes. These hybrid modes are excited by the hot electrons in Fe and then couple to vibrations in MgO in a mode-selective, non-thermal manner. After several picoseconds, the electrons and a sub-set of the phonons are in mutual equilibrium. A second slower energy transfer process by acoustic phonons contributes to thermalization of the entire heterostructure. Finally, after $90$\,ps all the different phonon sub-systems are equilibrated and the complete heterostructure is close to its fully thermalized state \cite{RothenbachFeMgO,Maldonado_PRB_17}. Due to the considerably larger specific heat of the lattice relative to the one of the electrons, the major part of the excess energy is then hosted by the phonon subsystems of the Fe- and MgO-constituent of the heterostructure. \\
In this paper, we report on the detailed inspection of the constituent-specific pump-induced spectral changes in transient soft X-ray spectroscopy of a Fe/MgO heterostructure upon laser excitation with different incident fluences in the range of $8-25$\,mJ/cm$^2$ on longer time scales (90\,ps) after thermalization. The analysis of the time-dependent spectroscopic fine structures at the O K-edge on the relevant time-scales with picosecond time resolution allows us to analyze the influence of thermal disorder very sensitively, as will be shown. For quantification of the effects, a simple simulation and fitting procedure of the pump-induced spectral features is introduced. We will demonstrate that the oxygen absorption K-edge of the insulator shows a uniform intensity decrease of the fine structures for an equilibrated state of the complete heterostructure. This individual pump-induced changes linearly depend on the incident laser fluence. Thereby, our work demonstrates the sensitivity of transient XAS to lattice excitations in oxides. The resulting lattice temperatures are quantified by comparison with ab initio calculations.

\begin{figure}
    \centering
    \includegraphics[width=0.99\columnwidth]{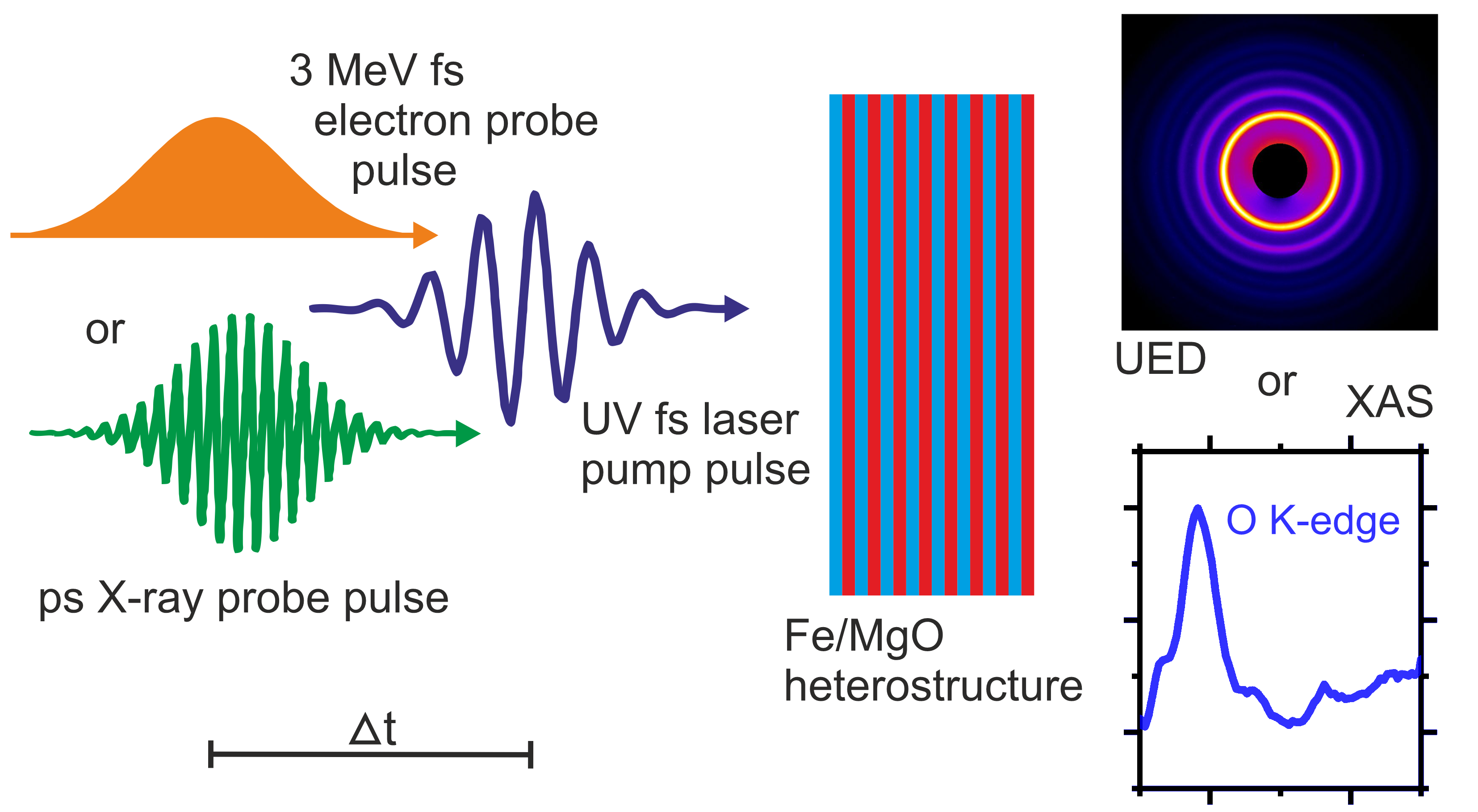}
    \caption{Schematic of the UV fs laser pump and ps soft X-ray absorption (XAS) or ultrafast electron diffraction (UED) probe experiment and the investigated [Fe/MgO] heterostructure sample. For more details see text.}
    \label{fig:fig1}
\end{figure}

\section{Samples and Methods}

The investigated [Fe/MgO]$_8$ heterostructures (sketched in Fig.~\ref{fig:fig1}) were grown by molecular beam epitaxy on a $200$\,nm thick Si$_3$N$_4$ membrane, which carries a $100$\,nm thick Cu heat sink on its backside in order to dissipate the excess energy deposited by the pump beam. The individual Fe and MgO layers are $2$\,nm thick, which was monitored during growth by a quartz-crystal balance and subsequently determined by X-ray diffraction. The sample preparation was carried out at a temperature of $400$\,K in a background pressure of $10^{-10}$\,mbar, excluding contamination of the individual deposited layers or oxidation of the Fe layers. Due to the use of the Si$_3$N$_4$ membrane as a substrate the Fe and MgO layer stacks are polycrystalline. The Fe-MgO interfaces are considered to be atomically sharp as confirmed by interface sensitive Conversion Electron M\"{o}ssbauer Spectroscopy, which showed that a potential intermixing of the constituents is limited to a maximum of one monolayer \cite{RothenbachFeMgO}.\\
The investigated Fe/MgO heterostructure shows a MgO band gap of $7.8$\,eV, with the Fermi-level of Fe located close to the midgap position (Ref.~\cite{RothenbachFeMgO}), resulting in an effective charge transfer gap of $\Delta$\,=\,$3.8$\,eV between occupied Fe states and the MgO conduction band, as reported in \cite{Petti_JoPCS_11}. Hence, laser excitation with a pump photon energy smaller than the bandgap of $7.8$\,eV provides a local pumping of exclusively the metal constituent. Although a hot electron mediated charge transfer upon absorption of an ultraviolet (UV) laser photon with an energy of $\Delta$\,$\leq$\,$h\nu$\,$<$\,$7.8$\,eV is energetically possible, we reported previously \cite{RothenbachFeMgO} that the electron-electron scattering in the Fe constituent suppresses this transfer effectively, so that it is not observed in the experiment. For more details on the excitation scheme of the Fe/MgO system, see Ref. \cite{RothenbachFeMgO} including the supplement. \\
Time-resolved element-specific X-ray absorption spectroscopy experiments were performed at the O K-edge to obtain a local, constituent specific probe of the insulator in the heterostructure. Experiments have been performed at the FemtoSpex facility of BESSY\,II (beamline UE56/1-ZPM) operated by Helmholtz Zentrum Berlin \cite{Holldack_JoSR_14}. We used  $50$\,fs, $266$\,nm laser pulses with an incident fluence ranging from $8$ to $25$\,mJ/cm$^2$ to induce transient spectral changes in the metal/insulator heterostructure. For the picosecond time-resolved experiments we used soft X-ray pulses with $70$\,ps duration, resulting in a total time resolution of $70$\,ps determined solely by the probe pulse. The energy resolution was $E/\Delta E$\,=\,$500$, resulting in $\Delta E$\,$\approx$\,$1.1$\,eV at the O K-edge ($\approx$\,$540$\,eV). The temporal overlap of the femtosecond UV pump and X-ray probe pulses has been determined by an independent transmission experiment through a $20$\,nm thick Fe reference film. \\

To directly study the lattice response of the heterostructures after fs laser excitation time-resolved ultrafast electron diffraction (UED) experiments have been performed at the MeV-UED facility at SLAC National Accelerator Laboratory \cite{weathersby15}. Experimental details have been presented in \cite{RothenbachFeMgO}. Relativistic ($E_{\rm kin} = 3.7$~MeV) electron pulses with 200\,fs duration (FWHM) and a bunch charge of approx.\ ${\rm 2\times 10^5}$ electrons, have been used to record Debye-Scherrer diffraction pattern of the polycrystalline samples after laser excitation in a normal-incidence transmission geometry over a large momentum transfer range of up to about 10~\AA$^{-1}$.

Samples for the UED experiments comprised similar [Fe/MgO]$_n$-heterostructures as for the XAS experiments, however with a reduced Si$_3$N$_4$ substrate thickness of 20\,nm and without an additional metal layer as heat sink. Both, a {\it symmetric} [2\,nm Fe / 2\,nm MgO]$_6$-, as well as an {\it asymmetric} [2\,nm Fe / 5\,nm MgO]$_5$-hetereostructure (to enhance the weak scattering from MgO) have been investigated. These samples were excited by 50\,fs laser pulses at wavelengths of 267\,nm and 400\,nm, corresponding to photon energies above and below the charge transfer gap, respectively, over an extended range of excitation fluences.

The X-ray absorption spectra of MgO were calculated in a first-principles way using the Korringa-Kohn-Rostoker (KKR) multiple scattering approach as implemented in the {\scshape spr-kkr} package \cite{SPRKKR_WWW,Ebert_RepProPhy_11}. The ground state spectrum was obtained within the atomic sphere approximation (ASA) using the generalized gradient approximation of Perdew, Burke and Ernzerhof \cite{Perdew_PRL_96} for the exchange-correlation functional and an angular momentum expansion up to f-states ($l_{max}$\,=\,$3$). We used a dense energy- (800 energy points along the contour line) and k-mesh (11368 points in the irreducible Brillouin zone, corresponding to a 55\,$\times$\,55\,$\times$\,55 mesh in the full zone) and took into account states up to $47$\,eV above the Fermi level. The finite temperature modification of O K-edge X-ray absorption spectra arising from lattice dynamics was modelled in the framework of the alloy analogy model \cite{Ebert_PRB_15} using the coherent potential approximation (CPA) to describe the average scattering problem for a set of thermally displaced ions with 14 independent displacements. Their magnitude was obtained according to the average square displacements within the Debye model with an appropriate value for the Debye temperature of MgO, $\Theta_D$\,=\,$743$\,K \cite{Debye_MgO}. 
Concentrating on the impact of the lattice vibrations on the spectra, we kept the volume of the two atom primitive cell constant at $a$\,=\,$4.214$\,{\AA}, thus neglecting the effect of thermally induced lattice expansion on the spectra, which one may consider of minor importance on the short time scales considered here. Finally, we applied to all spectra a broadening with a width of $1.5$\,eV to simulate the finite life-time of the excited states due to electron-electron interaction and the experimental energy resolution.

\section{Results}
\label{chap:results}

\begin{figure}
    \centering
    \includegraphics[width=0.99\columnwidth]{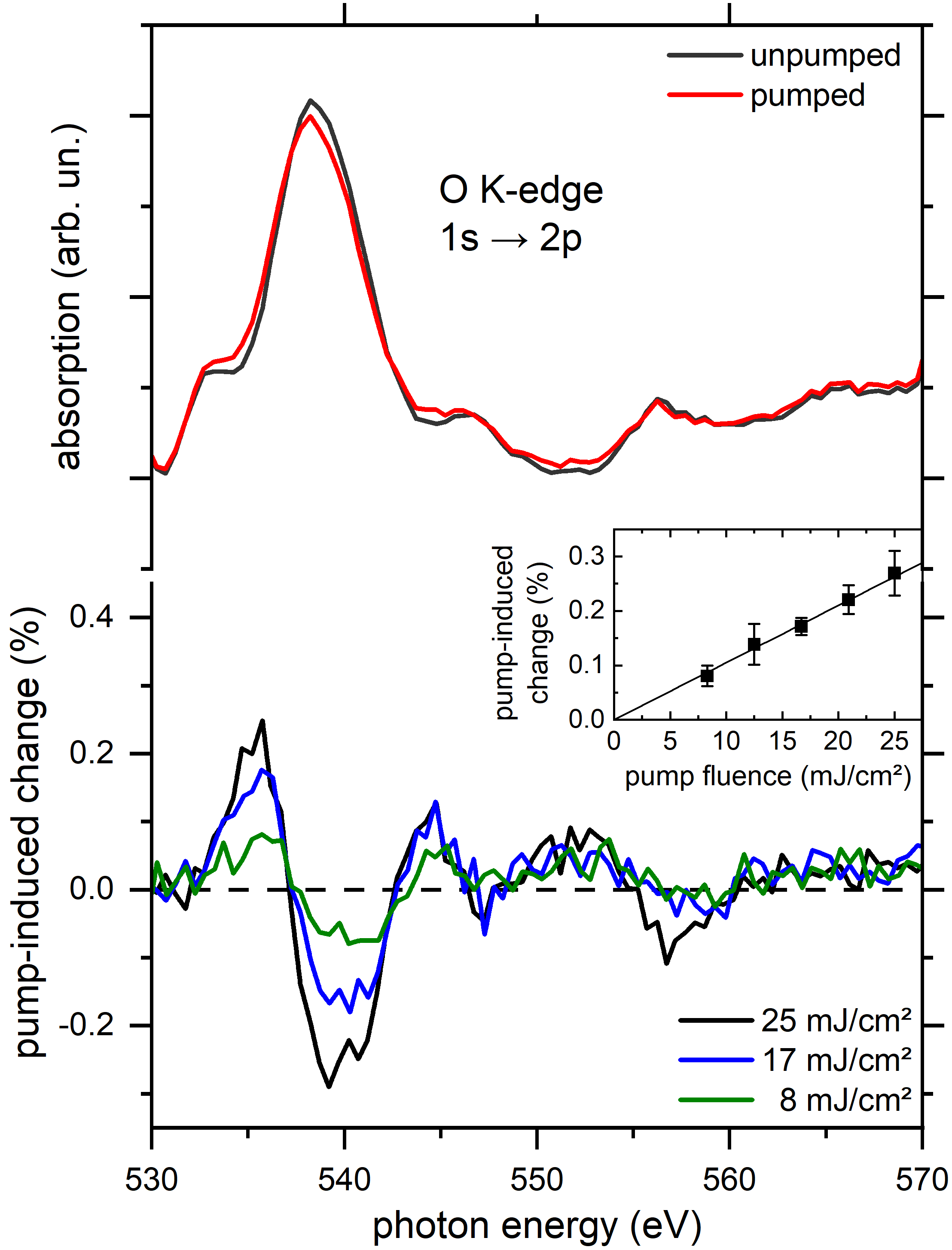}
    \caption{Top: X-ray absorption coefficient $\mu(E)$ at the oxygen K-edge before and after the UV-pump-pulse excitation (pump-probe delay time $90$\,ps, pump fluence 25\,mJ/cm$^2$). Bottom: Pump fluence dependence of the relative pump-induced change ($(\mu_{pumped}(E)-\mu_{unpumped}(E))/\mu_{unpumped}(E)$). The inset shows a linear dependence of the size of the pump-induced change, defined as half of the peak-to-peak value, on the pump fluence.}
    \label{fig:fig2}
\end{figure}

\subsection{O K-edge soft X-ray absorption spectroscopy}

Fig.~\ref{fig:fig2} shows the pump-induced changes at the O K-edge measured after excitation with $50$\,fs laser pulses with $266$\,nm wavelength at a fixed pump-probe delay of $90$\,ps for various laser pump fluences, ranging from $8$ to $25$\,mJ/cm$^2$. The size of pump-induced change, defined here as half of the peak-to-peak amplitude, shows a linear relation with the pump fluence (Fig.~\ref{fig:fig2}, inset). The pump-induced changes exhibit the same spectral shape for all fluences, originating from a constant intensity change of the fine structures of the X-ray absorption signal, measured before excitation with the laser pump pulse (so-called  unpumped spectrum) as it will be demonstrated below. It is one goal of this work to achieve a detailed understanding of the pronounced spectroscopic fine structures of the pump-induced signal. Obviously the pump-induced signal extends over an energy range of at least 40 eV. In the following section an analysis procedure is presented to simulate these changes in the spectral fine structure in this broader energy range.

\subsection{Simulation of the pump-induced change by intensity modification of the oxygen K-edge XAS}

\begin{figure}
    \centering
    \includegraphics[width=0.99\columnwidth]{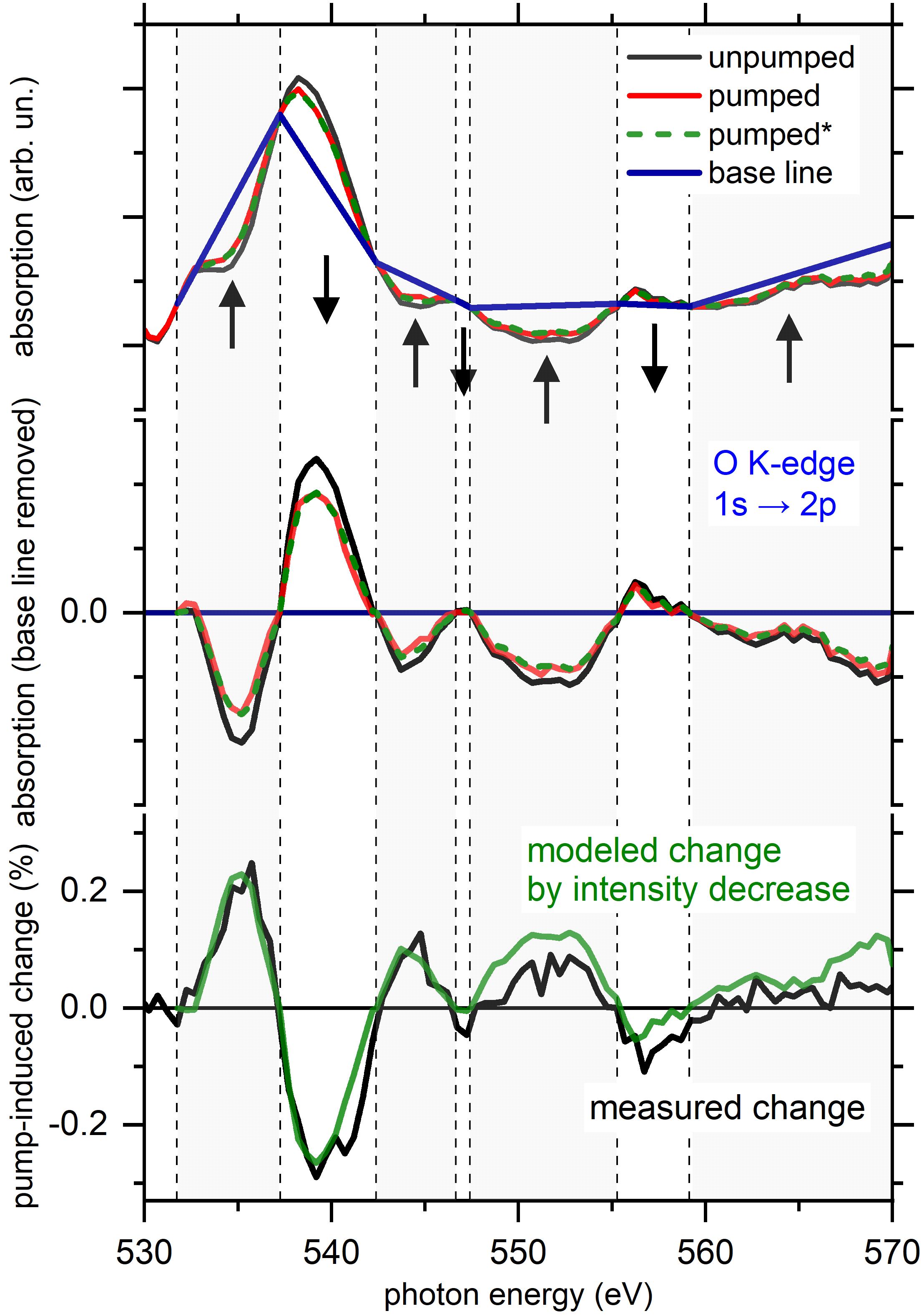}
    \caption{Depiction of the simulation and fitting procedure of the intensity decrease for the case of the O K-edge. Top:  Measured XAS at the oxygen K-edge before and after optical excitation (pump-probe delay time $90$\,ps, pump fluence $25$\,mJ/cm$^2$). Measured spectra are shown together with a linear base line (blue solid line) and an artificially generated pumped spectrum (green dashed line) by an amplitude reduction factor. Middle: Spectra of the top half after subtracting the base line. The vertical lines indicate the crossing points of the unpumped and pumped  spectra. The crossing points define distinct energy windows where the pumped spectrum has either lower or higher intensity than the unpumped spectrum in this region, indicated by the arrows. Bottom: Simulation of the pump-induced signal by an amplitude reduction factor compared to the experimental results. For details of the simulation see main text.}
    \label{fig:fig3}
\end{figure}

To simulate the pump-induced change by intensity modification, we analyze the measured soft X-ray absorption spectra as a function of photon energy for the O K-edge before (unpumped spectrum) and after optical excitation (pumped spectrum) as shown in the top half of Fig.~\ref{fig:fig3}, exemplarily for an incident pump fluence of $25$\,mJ/cm$^2$. It is clearly visible that there are definite crossing points of the unpumped and pumped spectra, which are marked by vertical lines. The crossing points define the limits of distinct energy windows where the pumped spectrum has either lower or higher intensity than the unpumped spectrum in this region as indicated by the arrows. These points are referred to as 'isosbestic points' in other contexts \cite{Yanong_JPCA,Smeigh_JACS}. The existence of the distinct energy windows is supported by theory, and will be discussed in the  section \ref{sec:theory}. By subtracting a linear base line (blue solid line), originating from linear interpolation of the crossing points, from the measured unpumped spectrum, the variation of the fine structure around the base line is determined as shown in Fig.~\ref{fig:fig3} in the middle. Multiplying these base line subtracted spectra with a constant reduction factor decreases the intensity of the fine structure variation (Fig.~\ref{fig:fig3}, middle) and, after adding the same base line, results in the simulated spectra, labeled as 'pumped$^{\ast}$' (Fig.~\ref{fig:fig3}, top). Indeed the measured pumped spectrum can be reproduced as can be seen in the agreement of the measured 'pumped' with the simulated 'pumped$^{\ast}$' spectra seen in Fig.~\ref{fig:fig3}. For a fluence of $25$\,mJ/cm$^2$ we identified an intensity decrease of the O K-edge with respect to the base line of $23$\,\% to model the experiment as accurate as possible. Please note that with this procedure only a single factor has to be applied to fit the intensity in the entire energy range of the measured data (see Fig.~\ref{fig:fig3}, bottom). 

The spectral changes we observe at the O K-edge upon laser excitation are in qualitative agreement with literature on LaCoO$_3$ \cite{Izquierdo_PRB_14}, where the corresponding O K-edge exhibits an intensity decrease of the fine structure. For LaCoO$_3$ it is found that at the  O K-edge, the spectral changes can be related to elevated lattice temperatures. \\
As shown in Fig.~\ref{fig:fig2}, we measured the pump-induced changes for various laser pump fluences, ranging from $8$ to $25$\,mJ/cm$^2$. The pump-induced changes exhibit the same spectral shape for all fluences. Hence, we can apply the introduced fitting procedure for all the measurements, and can indeed successfully describe all of our measured spectra with the individual simulation and fitting procedures. We find that the size of intensity decrease of the O K-edge fine structure depends linearly on the pump-fluence, just like the measured size of pump-induced change. We identify the largest effect of the intensity decrease with respect to the base line of $23$\,\% for the highest used fluence of $25$\,mJ/cm$^2$. For the lowest used fluence of $8$\,mJ/cm$^2$ we find the smallest effect of the intensity decrease of $7$\,\%. Table~\ref{tab:tabS1} summarizes the resulting dependence of the intensity decrease with respect to the base line on the pump-fluence. The temperature changes shown in this table are determined from density functional theory as explained below. The determination of the lattice temperature is a crucial step. Therefore, we carried out ultrafast electron diffraction experiments to analyze this quantity experimentally as shown in the next section. 

\begin{table}
\centering
\begin{tabular}{c c c c c}
\hline                        
pump fluence && intensity decrease&&  temperature change\\
 && &&  (from theory)\\
\hline\hline          
~\,8\,mJ/cm$^2$ && ~\,7\,\% &&  140\,K  \\
13\,mJ/cm$^2$  && 13\,\% && 250\,K  \\
17\,mJ/cm$^2$  && 14\,\% && 280\,K  \\
21\,mJ/cm$^2$  && 15\,\% && 300\,K  \\
25\,mJ/cm$^2$ && 23\,\% && 450\,K  \\ 
\hline           \\
\end{tabular}
\caption{Assignment of the observed intensity decrease (with respect to the base line) at the O K-edge for the measured pump fluences, to a calculated absolute temperature change.}
\label{tab:tabS1}
\end{table}

\subsection{Analysis of lattice temperature by ultrafast electron diffraction}

As discussed in detail in Ref.  [\onlinecite{RothenbachFeMgO}] in the ultrafast electron diffraction (UED) experiments the observed transient changes of the elastic scattering intensity can be fully attributed to an incoherent excitation of the lattice, i.e.\ changes of the r.m.s.\ atomic displacement ${\sqrt{\Delta\langle u^2\rangle}}$. These lead to an order-dependent decrease of the intensity of the different diffraction-peaks according to the Debye-Waller equation ($G_{hkl}$: length of the reciprocal lattice vector corresponding to a diffraction peak with Miller-indices h,k,l):
\begin{equation}
\label{eq:DWF}
\frac{I_{hkl}\left(t\right)}{I^0_{hkl}}=e^{-\frac{1}{3}\Delta\langle u^2\rangle (t) \cdot G^2_{hkl}}
\end{equation}

\begin{figure}
    \centering
    \includegraphics[width=0.99\columnwidth]{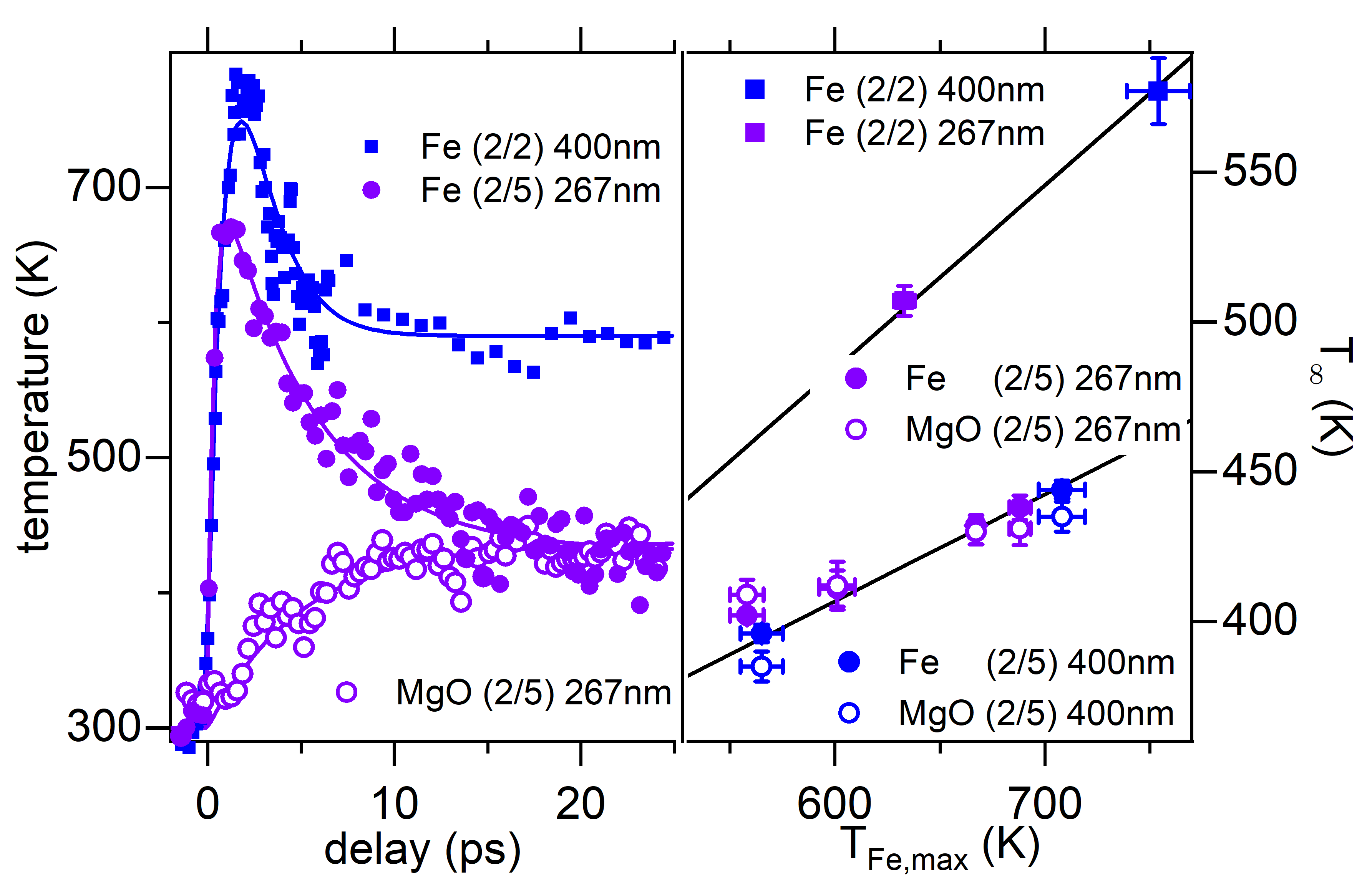}
    \caption{Time-resolved ultrafast electron diffraction to determine the temperature in the Fe/MgO multilayer systems after laser excitation. Left: time-dependence of the temperature in the Fe and MgO layers for symmetric (Fe(2\,nm)/MgO(2\,nm) labeled as (2/2)) and asymmetric multilayers (Fe(2\,nm)/MgO(5\,nm) labeled as (2/5)). The wavelength in nm of the pump pulse is given additionally. The labels 'Fe' or 'MgO' refer to the component-specific diffraction pattern of the heterostructure. Right: correlation of the asymptotic temperature $T_{asy}$ on long time scales with the maximum temperature in the Fe layers $T_{Fe,max}$.}
    \label{fig:fig4}
\end{figure}

Using published temperature dependent Debye-Waller factors for Fe (e.g., Ref. \onlinecite{kharoo77} and references therein) and MgO \cite{baldwin64, singh82} the measured $\Delta\langle u^2\rangle (t)$ are converted into time-dependencies of the lattice temperature of both constituents of the heterostructures. Typical results are shown in Fig.\ \ref{fig:fig4} (left) for the {\it asymmetric} heterostructure after 267\,nm excitation with a fluence of $F$ = 9\,mJ/cm$^2$ (violet data points; filled circles: Fe; open circles: MgO) as well as the {\it symmetric} heterostructure after 400\,nm excitation with $F$ = 5\,mJ/cm$^2$ (blue data points: Fe). Here it should be stressed that it is not straight forward to compare the pump pulse fluences in the ultrafast electron diffraction measurements with the X-ray absorption experiments as the absorbed fluence is crucial whereas the incident fluence is given. However, both fluences are in the same order of magnitude, with the ones in the UED being presumably all slightly smaller than in the X-ray absorption spectroscopy.

Before discussing these results, two points need to be emphasized. First, the changes of the diffraction pattern are dominated by Fe, in particular at early delay times. Therefore, and due to the fact the diffraction peaks of Fe and MgO partially overlap, the lattice temperature in MgO could be reliably determined only for longer delay times (i.e.\ after a few ps) and only for the {\it asymmetric} heterostructure due to its increased MgO-content. Second, the temperatures determined at early delay times (up to 2\,ps) must be interpreted with some care since the phonon systems exhibit initially non-thermal populations \cite{chase16, RothenbachFeMgO, maldonado20}. However, since our analysis focuses on the relaxation and equilibration of the heterostructures and thus on the behavior at later times, this is a less serious problem.

For both heterostructures the lattice temperature in Fe rises rapidly, reaching its maximum value after about 1 - 2\,ps, indicative of the strong electron-phonon interaction in Fe, which is further enhanced by coupling of excited electrons in Fe to hybrid interface vibrational modes as reported in \cite{RothenbachFeMgO}. Subsequently the temperature in Fe starts to decrease reaching a stationary value after about 15 - 20\,ps (described here as the {\it asymptotic} temperature), while the temperature in MgO increases towards the same value as in Fe. The {\it asymptotic} temperature values in Fe and MgO have been determined using Debye-Waller factors for an equilibrium situation. That these values for Fe and MgO coincide, clearly indicates that the whole heterostructure has reached an equilibrium state and that a thermal description is justified at these later time delays.

This is also demonstrated by the data presented in Fig.\ \ref{fig:fig4} (right), which shows the {\it asymptotic} temperature of Fe (filled symbols) and MgO (open circles) as a function of the maximum temperature in Fe (corresponding to data obtained at different laser fluences, blue and violet data points correspond to excitation with 400\,nm and 267\,nm laser pulses, respectively). For the {\it asymmetric} heterostructure, where a reliable temperature determination in MgO had been possible, the {\it asymptotic} temperatures in Fe and MgO are identical (within the errors of the measurements) over the whole range of excitation fluence and follow a linear dependence (starting at 300\,K / 300\,K). A similar dependence is observed for the {\it symmetric} heterostructure. However, due to the smaller MgO-thickness the asymptotic temperature (which could here only determined for Fe) is higher at a given maximum temperature than for the {\it asymmetric} heterostructure. 

In conclusion, the UED-measurements demonstrate that the heterostructures fully equilibrate within 15 - 20\,ps and that for a {\it symmetric} structure the temperature in MgO reaches values of 500 - 550\,K for laser fluences around 10\,mJ/cm$^2$.

\subsection{DFT calculations}
 \label{sec:theory}

\begin{figure}
    \centering
    \includegraphics[width=0.99\columnwidth]{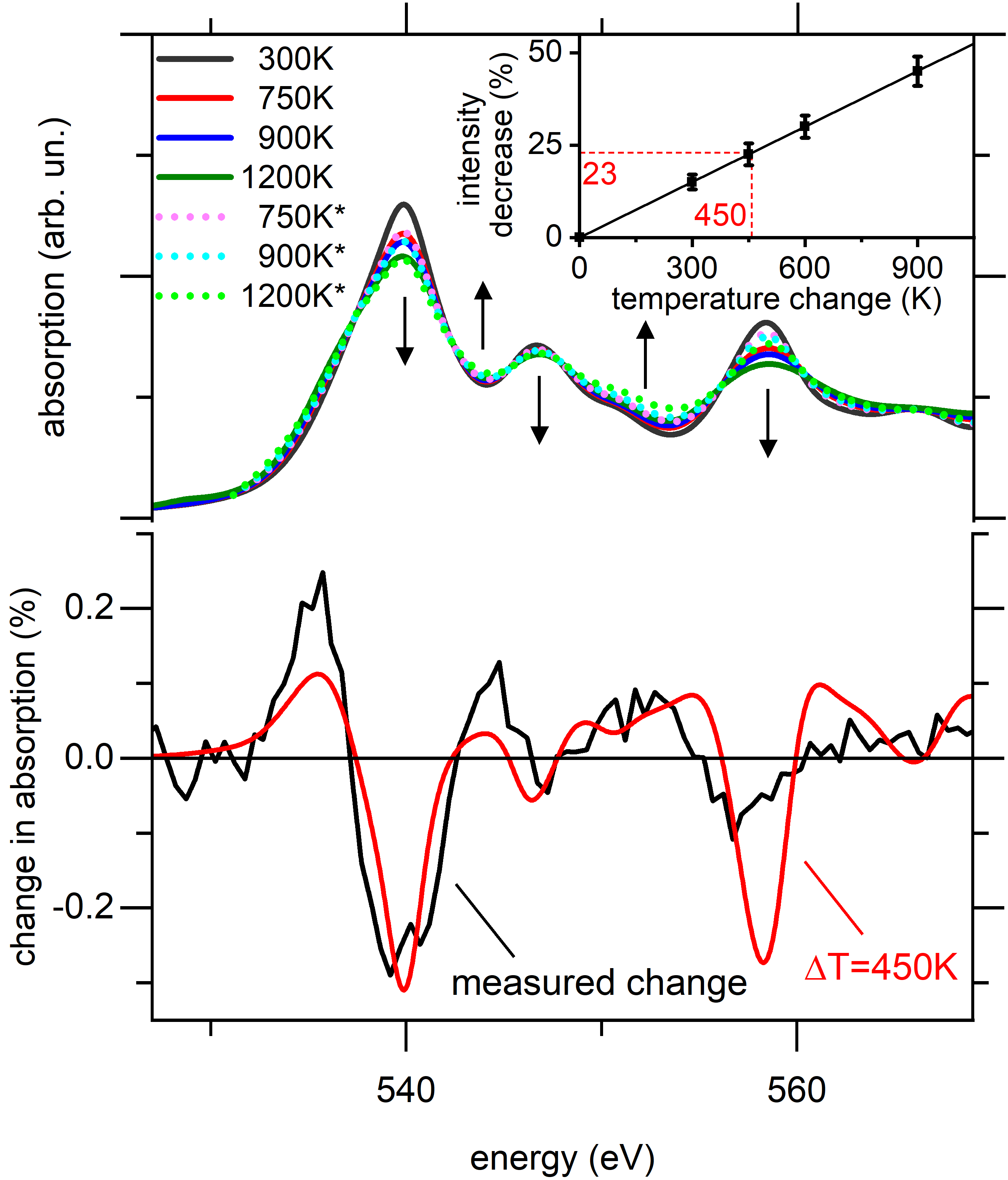}
    \caption{Top: DFT calculations of the O K-edge XAS of bulk MgO at four different temperatures modelled with a finite set of thermal displacements in the framework of the alloy analogy model \cite{Ebert_PRB_15,Ebert_Springer_18} (solid lines). With increasing temperature, the calculation shows an overall intensity suppression of the fine structures. There are well defined energy regions where the XAS at elevated temperature has lower or higher energy than the spectrum at lower temperature, indicated by the arrows. Additionally, spectra generated from the $300$\,K spectrum (dotted lines) are shown (modeling according to Fig.~\ref{fig:fig3}). These match the corresponding finite temperature KKR-CPA spectra. Bottom: Comparison of the experimental pump-induced signal (Fig.~\ref{fig:fig2}, pump fluence 25\,mJ/cm$^2$) with the relative difference of the calculated XAS at 300\,K and 750\,K. For this comparison relative changes after subtraction of a base line (see Fig.~\ref{fig:fig3}) are considered.}
    \label{fig:fig6}
\end{figure}

Fig.~\ref{fig:fig6} shows results of DFT calculations of the O K-edge XAS of bulk MgO using the KKR-CPA approach. The calculations were carried out with the alloy analogy model \cite{Ebert_PRB_15,Ebert_Springer_18} for different lattice temperatures. 
Naturally the DFT approach applies to larger timescales beyond $20$\,ps. Here, the phonon systems of the individual layers and the entire heterostructure have equilibrated, making a quasi-static description by a common lattice temperature feasible. Moreover, the equilibrated state of the heterostructure allows the comparison of the spectra of the O K-edge in experiment (heterostructure) and theory (bulk MgO). These results support our interpretation of the experimental data. Changes in the vibrational density of states of MgO due to hybridization with Fe phonon modes at the interface occur for high energy phonon modes in the heterostructure. However, their influence is mainly restricted to the non-thermal, sub-ps energy transfer dynamics \cite{RothenbachFeMgO}. On the longer timescales analyzed here, the lower energy acoustic phonon modes play a dominant role. Matching with our experimental results, the calculation shows that with increasing temperature, there is essentially no shifting of the spectral feature at the O K-edge but that the overall fine structures are suppressed in their intensity (Fig.~\ref{fig:fig6}). Moreover, the calculation demonstrates that there are indeed defined regions where the XAS at elevated temperature has lower or higher intensity than the spectra at lower temperature, thus supporting our simulation procedure (see Fig.~\ref{fig:fig3}). The energy regions defined in our experimental analysis seen  in Fig.~\ref{fig:fig3} are nearly identical in the DFT calculations as can be seen in the very similar zero crossings in Fig.~\ref{fig:fig6} (bottom). \\
Consequently, we can relate the magnitude of the intensity suppression to an induced lattice temperature change: Using the spectrum at $300$\,K, we generate the spectra at elevated temperatures by the same procedure. Fig.~\ref{fig:fig6}, shows that the simulated spectra (dotted lines) match the calculated ones (solid lines) in intensity and shape reasonably well. With this relation we can show that the combination of our simulation and fitting procedure and the KKR-CPA calculations offer a powerful possibility to analyze and quantify measured pump-induced lattice temperature changes in an insulator. \\
The intensity decrease of the O K-edge with respect to the base line of $23$\,\%, measured for a fluence of $25$\,mJ/cm$^2$ resembles a lattice temperature change of $450$\,K. This value nicely connects to the temperature increase determined by the ultrafast electron diffraction experiments. Furthermore, this temperature change is of the same order of magnitude as previous findings of an equilibration temperature of the heterostructure at pump-probe delays of $>5$\,ps, as determined by ultrafast electron diffraction measurements \cite{RothenbachFeMgO}, where it was shown that a fluences of only $9$\,mJ/cm$^2$ yielded  a temperature change of $\Delta T_{\mathrm{MgO}}$\,=\,$(130 \pm 10)$\,K at time delays larger than 5\,ps. These results match with the results presented in here in Table~\ref{tab:tabS1} for lower fluences.\\

\subsection{Discussion}

In a simplified picture one can view the effect of the lattice vibrations on the near edge spectra in the following way: The XAS of the sample at low temperature (T=0\,K in the DFT calculation) can be viewed as a reference spectrum. This spectrum exhibits pronounced fine structures. Increase of temperature leads to a broadening of the fine structures with the consequence that peaks are reduced and valleys are filled. In the extended energy range of the XAS the oscillatory fine structures are denoted as extended X-ray absorption fine structure (EXAFS). These fine structures originate from constructive and destructive interference phenomena of the outgoing photoelectron wave being scattered at the surrounding atoms. Lattice excitations lead to a reduction of the oscillatory fine structure. In the present work we focus on the near edge regime which can be described in a mutiple-scattering approach. In the following we discuss the suppression of the O K-edge fine structures in the near edge region observed in the experiment (Fig.~\ref{fig:fig2}) and theory (Fig.~\ref{fig:fig6}) with increasing temperature. We want to draw a connection to the extended energy regime and the effect of the increase of thermal disorder. In the extended energy regime the EXAFS Debye-Waller factor $e^{-2\cdot \sigma^2(T)\cdot k^2}$, with $k$ being the the wave number of the photoelectron, describes the reduction of the oscillatory fine structures with increasing temperature $T$ by the increasing mean square relative displacement (MSRD) $\sigma^2(T)$. The value of $\sigma^2(T)$ can differ from $\langle u^2\rangle$ given in equation \ref{eq:DWF} as the relative motion of absorbing and backscattering atom is relevant in the EXAFS process whereas the absolute values of the mean square displacements are crucial for the amplitude reduction in diffraction experiments. An approximative way to account for atomic vibrations in the XAS near-edge region by extending this method was suggested and applied in the past 
\cite{Fujikawa_JPSJ,Sipr_JSR}. It consists in multiplying the free-electron propagator in the multiple-scattering series by the Debye-Waller factor, meaning that only pair-wise fluctuations of interatomic distances are considered. As the Debye-Waller factor reduces both, the positive as well as negative part of the EXAFS oscillations, this procedure would lead to reducing the peaks and filling the valleys in the near-edge fine structure. It follows from Figs.~\ref{fig:fig3} and \ref{fig:fig6} that this simple picture explains the trends visible in the experiment as well as in more advanced calculations. Because of the analogy in describing the thermal disorder in the EXAFS region and in the near edge region, one can understand that elevated phononic temperatures do not induce an energy shift but an intensity decrease of the near edge fine structures.

By comparing the amplitude of the simulated spectral changes to the calculated spectra for elevated lattice temperatures (cf. Fig.~\ref{fig:fig6}), we are able to assign the observed intensity decrease at the O K-edge to an absolute temperature change. Table~\ref{tab:tabS1} summarizes the resulting relation between the pump-fluence, the intensity decrease with respect to the base line and the temperature change for all of our measurements. The experimentally determined amplitude reduction of 23\,\% is connected to a temperature increase of $\Delta T$=450\,K as deduced from the DFT calculations. This temperature increase is in the same order of magnitude as determined by the ultrafast electron diffraction experiments for the symmetric multilayer for similar pump fluences presented in Fig.~\ref{fig:fig4}(left) and justifies our analysis.

\section{Conclusion}

In conclusion, we report the spectral changes of the oxygen K-edge XAS of a metal/insulator heterostructure upon laser excitation. For the oxygen K-edge of the insulator an elevated phononic temperature manifests in a uniform intensity decrease of the fine structure. In order to quantify these changes, the spectral signatures can be simulated and fitted by a simple procedure which considers an intensity decrease of the XAS fine structures. This interpretation is further supported by comparison to DFT-based calculations of the temperature-dependent spectra. For all incident laser fluences of up to $25$\,mJ/cm$^2$ the pump-induced changes show a linear dependency on the pump fluence. Our work demonstrates the sensitivity of transient near edge XAS to phonons. Similar modeling might in the future be exploited to analyze and quantify the nature of lattice excitations also under conditions of phonon non-equilibrium, where a description by a common lattice temperature is not applicable. \\

\begin{acknowledgments} 
We acknowledge fruitful discussions with J.~J.~Rehr and E.~K.~U.~Gross. This work is financially supported by the Deutsche Forschungsgemeinschaft (DFG, German Research Foundation) through the Collaborative Research Centre (CRC) 1242 (project number 278162697). Calculations were carried out on the MagnitUDE supercomputer system (DFG grants INST 20876/209-1 FUGG, INST 20876/243-1 FUGG) of the Centre for Computational Sciences and Simulation at the University of Duisburg-Essen. Helmholtz Zentrum Berlin is gratefully acknowledged for the allocation of synchrotron radiation beamtime and for financial support for travel to BESSY II. UED@SLAC is supported in part from the U.S. Department of Energy BES SUF Division Accelerator \& Detector R\&D program under contract No.'s DE-AC02-05-CH11231 and DE-AC02-76SF00515. 
We thank N.~Pontius, R.~Mitzner, K.~Holldack, C.~Sch\"{u}{\ss}ler-Langeheine and 
E.~W\"{u}st for experimental support and U.~von~H\"{o}rsten for his expert technical assistance with sample preparation.
\end{acknowledgments}

\bibliography{rothenbach_PRB.bib}

\end{document}